\documentclass[superscriptaddress,showkeys,twocolumn,nofootinbib,longbibliography]{revtex4-1}

\usepackage{amsmath}
\usepackage{amssymb}
\usepackage{graphicx}
\usepackage{epsf}
\usepackage{slashed}
\usepackage{enumitem}
\usepackage[czech,english]{babel}
\usepackage[cp1250]{inputenc}
\usepackage{hyperref}
\usepackage{xcolor}
\usepackage{physics}

\usepackage[only,llbracket,rrbracket,llparenthesis,rrparenthesis]{stmaryrd} 
\usepackage{accsupp} 

\usepackage[font=small,labelfont=bf,justification=raggedright]{caption}
\usepackage{subcaption}

\renewcommand{\Re}{\mathop{\rm Re}\nolimits}

\usepackage{bbm} 

\usepackage[nodayofweek]{datetime}
\newdateformat{mydate}{\twodigit{\THEDAY}{ }\shortmonthname[\THEMONTH], \THEYEAR}
\usepackage[normalem]{ulem}

\usepackage{color}
\usepackage{ulem} 

\definecolor{myred}{rgb}{1,0,0}
\definecolor{mygreen}{rgb}{0,0.8,0.2}
\definecolor{myblue}{rgb}{0,0,1}

\definecolor{Ared}{rgb}{1,0.7,0}
\definecolor{Agreen}{rgb}{0.7,0.8,0.2}
\definecolor{Ablue}{rgb}{0,0.7,1}

\renewcommand{\emph}[1]{\textit{#1}}

\begin{document}


\title{Resonance phenomena in vortex-antivortex collisions}

\author{M. Bachmaier}

\affiliation{Arnold Sommerfeld Center, Ludwig-Maximilians-Universitat, Theresienstrasse 37, 80333 Munchen, Germany}
\affiliation{Max-Planck-Institut fur Physik, Boltzmannstrasse 8, 85748 Garching, Germany}

\author{A. Wereszczynski}

\affiliation{Institute of Theoretical Physics, Jagiellonian University,
Lojasiewicza 11, Krak\'{o}w, Poland}
\affiliation{International Institute for Sustainability with Knotted Chiral Meta Matter (WPI-SKCM$^{\; 2}$), Hiroshima University, 1-3-1 Kagamiyama, Higashi-Hiroshima, Hiroshima 739-8526, JAPAN}

\begin{abstract}

In this work, we provide a full map of scattering scenarios between a Nielsen-Olesen vortex and antivortex. Importantly, in the deep type II regime, such a collision reveals a chaotic pattern in the final state formation with bounce windows immersed into annihilation regions. This structure is due to the energy transfer mechanism triggered by a {\it quasinormal mode}, specifically the {\it Feshbach resonant mode}, hosted by the vortex. 
\end{abstract}

\maketitle

\section{Motivation}
The solitons-antisoliton collision is a fundamental process that governs the dynamics of the ensemble of solitons produced in a phase transition. In the simplest case of domain walls (kinks), it leads to the appearance of the famous chaotic self-similar pattern in the final state formation~\cite{Sugiyama:1979mi, Campbell:1983xu}. Concretely, the pair of the kink and antikink can annihilate to the vacuum, via the formation of a long-lived oscillating state called the oscillon (or bion) or can be backscattered after a finite number of collisions ({\it bounces}). The actual scenario is very sensitive to the initial conditions, e.g., the initial velocity $v_{\rm in}$, revealing a fractal-like structure. In particular, between the annihilation (smaller $v_{\rm in}$) and the one-bounce backscattering (bigger $v_{\rm in}$), one finds a self-similar pattern of multi-bounce and annihilation regions, known as the {\it bounce windows} and {\it bion chimneys}, respectively. 

This behavior is explained by the resonant energy transfer mechanism~\cite{Sugiyama:1979mi, Campbell:1983xu, Manton:2021ipk}. During the collision, the energy flows between the kinetic and internal degrees of freedom (DoF). If too much energy is stored in the internal DoF, the soliton becomes unable to overcome the attractive kink-antikink interaction, resulting in another collision. However, in certain circumstances, the energy may flow back to the kinetic DoF, allowing the kinks to reappear as the final states. Obviously, the necessary condition for such a process is the existence of an internal mode. Typically, it is a massive bound mode of the single kink, but it can also be a quasinormal mode~\cite{Dorey:2017dsn, GarciaMartin-Caro:2025zkc} or a mode of the kink-antikink configuration~\cite{Dorey:2011yw, Adam:2022mmm}.

Since physics in one dimension often shows phenomena unique to that setting, one can ask whether collisions of higher dimensional solitons lead to a similar pattern. Here, the vortices in the Abelian-Higgs (AH) model can serve as a prototypical example. 
In fact, it has recently been understood that the dynamics of multiple Nielsen-Olesen vortices is very sensitive to the excitation of the internal modes. The scattering of the excited BPS vortices (optionally equipped with a force due to the static interaction of non-BPS vortices) exhibit dynamics that go well beyond the geodesic flow on the moduli space\footnote{Beyond soliton dynamics, we also refer to a recent discussion on how perturbations of black holes can modify their dynamics through the so-called ``swift memory burden effect"~\cite{Dvali:2025sog}.}.
The reason is the appearance of the mode generated force. This force is able to change the usual right-angle scattering in the head-on collisions of two BPS vortices~\cite{Ruback:1988ba, Shellard:1988zx, Myers:1991yh, Samols:1991ne, Stuart:1994yc} into various scenarios depending on which mode is excited. For example, an excitation of the lowest $2$-vortex mode gives rise to multi-bounce right-angle scatterings where the actual number of bounces changes with the initial velocity and phase of the vortices in a chaotic, and probably fractal manner~\cite{Krusch:2024vuy, AlonsoIzquierdo:2024nbn, Alonso-Izquierdo:2025suz, Alonso-Izquierdo:2024fpw}. As in the case of the kinks, the multi-bounce scattering is an effect of the resonant energy transfer mechanism, where, during the collision, the kinetic energy of the vortices (excitation of the zero modes) can be temporarily transferred into potential energy of the vibration modes~\cite{AlonsoIzquierdo:2024nbn}. This increases the attraction due to the mode generated force and the vortices have too little kinetic energy to overcome the attraction and thus bounce once again. Sometimes, during the collision, the energy can be transferred back to the kinetic degrees of freedom and the vortices separate and appear as the final states. 

\vspace*{0.2cm}

In the present work, we focus on the vortex-antivortex scattering in the Abelian-Higgs model at various values of the coupling parameter $\lambda$. The aim is to verify to what extent the resonant energy transfer mechanism governs the soliton-antisoliton scatterings in higher dimensions. Such collisions have been briefly analyzed before. Vortices have been shown to annihilate, except for a very high collision velocity, where they can recreate themselves as free final states \cite{Myers:1991yh}. We will show that the pattern of the collision is much more fascinating.

\section{The Abelian-Higgs Model}
\begin{figure*}
    \centering
    \includegraphics[trim=0 0 0 25,clip,width=0.9\textwidth]{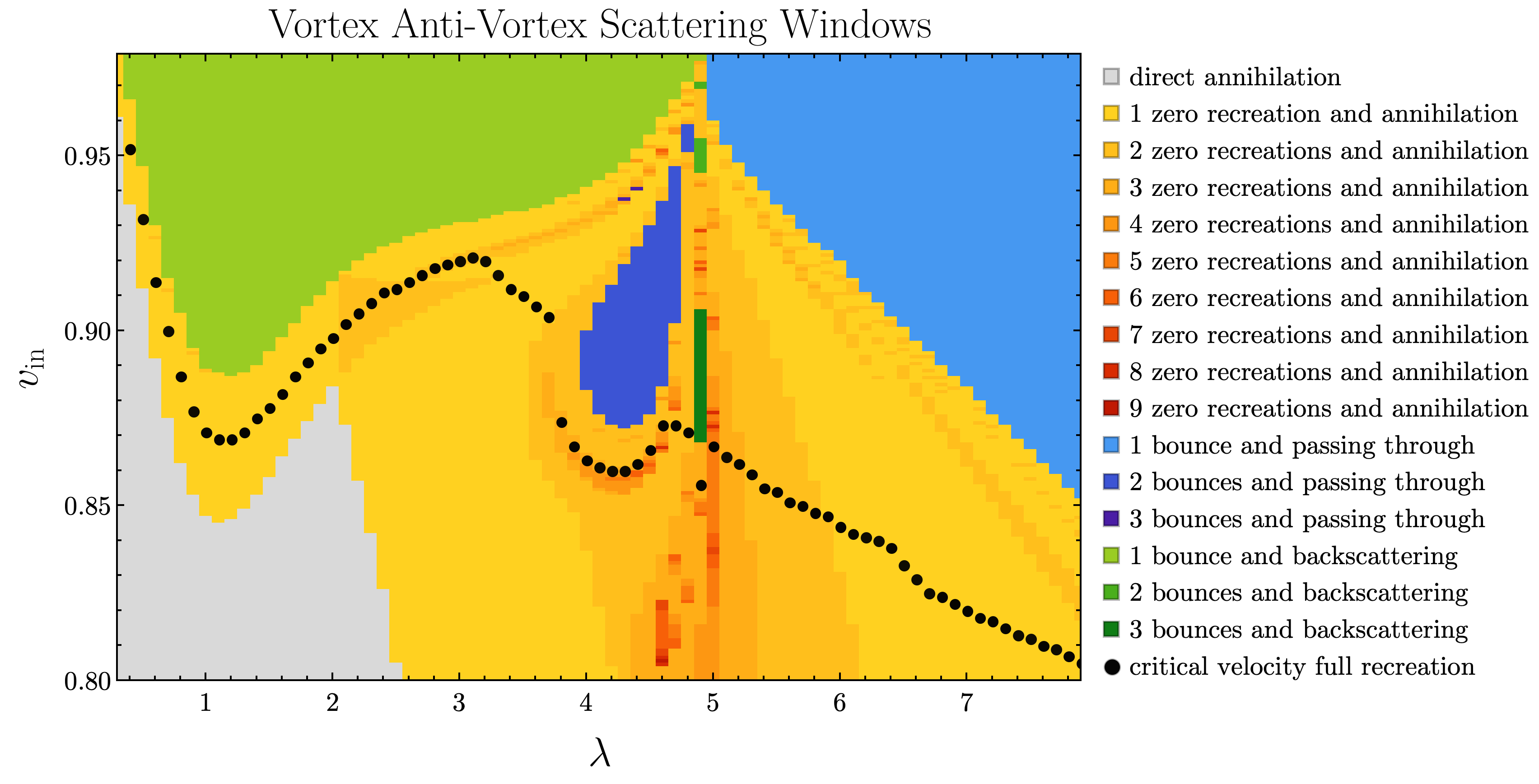}
    \caption{A summary of the different outcomes in vortex-antivortex collisions. On the $x$-axis, the values for $\lambda$ are given. On the $y$-axis, the initial velocity is given.}
    \label{fig:chaoticmap}
\end{figure*}

We will consider the Abelian-Higgs model in $2+1$ dimensions, which is defined by the following Lagrangian density
\begin{equation}
\mathcal{L}=-\frac{1}{4} F_{\mu\nu}F^{\mu \nu} +
\frac{1}{2} \overline{D_\mu \phi}\, D^\mu \phi -\frac{\lambda}{8}
(\overline{\phi}\, \phi-V^2)^2,
\label{action1}
\end{equation}
where $D_\mu \phi = (\partial_\mu -ig
A_\mu )\phi$ is the covariant derivative of the complex scalar field $\phi$ and  $F_{\mu\nu}=\partial_\mu A_\nu - \partial_\nu
A_\mu$ is the electromagnetic field tensor for the gauge field $A_\mu$.
If we rescale to dimensionless coordinates $x^\mu\mapsto \frac{x^\mu}{gV}$, the full theory depends on a single parameter, the Higgs mass $m_h=\sqrt{\lambda}V$.
Therefore, in the following, we set $g=1$ and $V=1$ without loss of generality, leaving the setup fully parametrized by $\lambda$. Furthermore, we will work in units of $c=1$.
The field equations read as follows 
\begin{align}
    D_\mu D^\mu \phi + \frac{\lambda}{2}\phi (|\phi|^2-1)&=0, \\
    \partial_\mu F^{\mu \nu} -\frac{i}{2}(\phi \partial^\nu \bar{\phi} - \bar{\phi} \partial^\nu \phi) +|\phi|^2A^\nu &=0.
\end{align}
This theory has a vortex (and antivortex) solution, which is given by~\cite{Nielsen:1973cs} 
\begin{equation}
\label{eq:vortex-ansatz}
    \phi = f(r)e^{\pm i\theta},\;\; A_\theta=\frac{a(r)}{r},\;\; A_r=0,
\end{equation}
where the profile functions $f(r)$, $a(r)$ obey the ordinary differential equations
\begin{align}
    \frac{\dd^2f}{\dd r^2}+\frac{1}{r} \frac{\dd f}{\dd r} - \frac{1}{r^2}f( \pm 1- a)^2+\frac{\lambda}{2} f(1-f^2) &= 0, \\
    \frac{\dd^2a}{\dd r^2}-\frac{1}{r} \frac{\dd a}{\dd r} +(\pm 1- a)f^2 &=0.
\end{align}
The topologically non-trivial boundary conditions are 
\begin{align}
    f(\infty) = 1,&\hspace{1cm} \; a(\infty)= \pm 1,\nonumber\\
    f(0)=0,&\hspace{1.19cm} \; a(0)=0.
\end{align}
A stable static unit-charge vortex exists for any value of the coupling constant $\lambda$. However, this is not the case for vortices of higher charge. Such (stable) solutions exist only for $\lambda < 1$ (type I vortices) and for $\lambda = 1$ (BPS vortices). This reflects the fact that unit-charge vortices attract in the type I regime or in the BPS case in which they experience no static force. For $\lambda > 1$ (type II vortices), there are no stable multi-vortex configurations, since the vortices repel each other~\cite{Speight:1996px, Speight:2025qrr}. 

Finally, for all $\lambda$, a vortex and an antivortex always attract each other.

\section{Vortex-antivortex collisions}

As an initial configuration for the simulation we take a vortex and an antivortex that are boosted toward each other with an initial velocity of $v_{\rm in}$.
The details of the numerical implementation are given in the Supplementary Material. In this work, we restrict our analysis to head-on collisions, implying that the problem has $y \to -y$ symmetry. All motion of the zeros of the Higgs field (position of the vortices) takes place solely along the $x$-axis. 

The interesting dynamics occurs in a highly relativistic regime ($v_{\rm in} > 0.8$).
Furthermore, we investigated a rather large range of the coupling constant, $\lambda \in [0.1,\, 8.0]$, which allowed us to probe the strong type I and type II regimes. In our numerical analysis, we identified several scenarios in the vortex-antivortex collisions. A brief summary of the subsequent discussion, together with the animated results of the numerical simulations, is provided in the following video: \url{https://youtu.be/1o__huMd13o}.

First, as previously observed~\cite{Myers:1991yh}, at a very high initial velocity $v_{\rm in}>v_{\rm cr}(\lambda)$, the solitons collide once and then are recreated in the final state. The recreation can result in two possible outcomes, consistent with the symmetry of the system. In the first case, the vortices backscatter, such that the vortex is entering from the left and leaving to the left. In the second case, they pass through one another, with a vortex entering from the left and leaving to the right. An illustrative explanation of our naming convention is provided in Figure~\ref{fig:naming} of the Supplemental Material. Interestingly, for $\lambda<\lambda_{\rm cr}\approx 5$, the vortices are backscattered, while for $\lambda>\lambda_{\rm cr}$ they pass through one another, see Figure \ref{fig:chaoticmap} (light blue and light green regions). These two regimes are separated at $\lambda=\lambda_{\rm cr}$, where the critical velocity, $v_{\rm cr}(\lambda_{\rm cr})$, has a local maximum.

The second main channel of the interaction is the annihilation. It dominates at smaller initial velocities, see Figure \ref{fig:chaoticmap} (gray, yellow, orange, and red regions). The annihilation can be instantaneous, that is, without the formation of any temporary state (gray region). For smaller $\lambda$, this is a dominating scenario that already occurs for quite large $v_{\rm in}$. However, the vortices can temporarily be recreated and perform several bounces (collisions). We identify the recreation as the appearance of two zeros of the scalar field. For larger $\lambda$ the recreation scatterings begin to dominate. Of course, the reappearance of the zeros not necessarily means the full recreation of the vortices.  In Figure~\ref{fig:chaoticmap}, we also show the velocity above which the vortices are fully recreated (black dots), such that the separation between the zeros exceeds the vortex size $d$. Here, $d/2$ is the radius enclosing $90\%$ of the static energy. Notice that for $\lambda \gtrsim 3.2$, the recreation velocity starts to decrease. This behavior can be attributed to the proximity of bounce windows triggered by Feshbach resonances, which we will discuss in section~\ref{sec:the-perturbation-of-the-vortex}.

These two scenarios fully describe the dynamics for $\lambda < 4.0$. For $v<v_{\rm cr}$ we find the annihilation
regime while for $v>v_{\rm cr}$ the vortices backscatter.

However, when $\lambda \gtrsim 4.0$, that is, deep in the type II region, the situation becomes much more involved. Here, we identified multi-bounce windows, which are the regions where the vortices reappear after the subsequent collisions but have enough kinetic energy to separate and form free independent objects in the final state. They are visible in Figure~\ref{fig:chaoticmap} as darker blue or darker green regions, where the color denotes whether the vortices are backscattered (green) or pass through each other (blue). In particular, the bounce windows seem to accumulate in the region where $\lambda \to \lambda_{\rm cr}$.

\begin{figure}
    \centering
    \includegraphics[trim=0 0 0 35,clip,width=\linewidth]{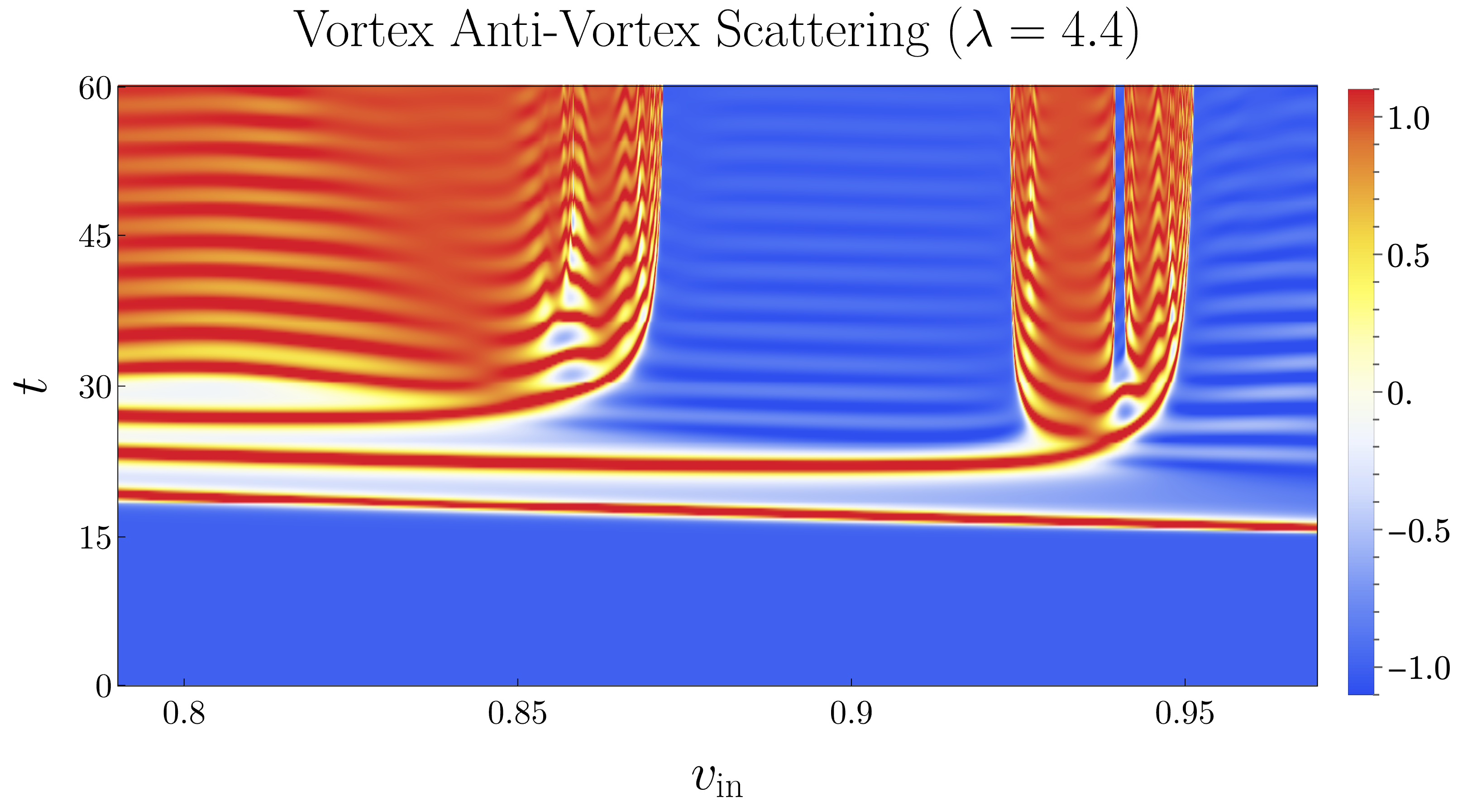}
    \caption{Time evolution of the real part of the scalar field at the origin, $\Re \phi(\vec{x}=0, t)$, during a vortex-antivortex scattering for $\lambda = 4.4$, shown for different initial velocities $v_{\rm in}$.}
    \label{fig:fractal-structure-lambda44}
\end{figure}

In Figure~\ref{fig:fractal-structure-lambda44}, we present the scatterings for $\lambda=4.4$. Here, we plot the real component of the scalar field at the origin as a function of time for various initial velocities $v_{\rm in}\in [0.79,\, 0.97]$. We found that there is a wide two-bounce window for $v_{\rm in} \in [0.8718,\, 0.9238]$ and a very narrow three-bounce window around $v_{\rm in}\in [0.9397,\, 0.9409]$. These windows are immersed in annihilation regions. In the Supplement Material we also show the final velocity of the outgoing vortex with respect to the initial velocity. 

\begin{figure*}
    \includegraphics[trim=0 66 80 30,clip,width=0.34\textwidth]{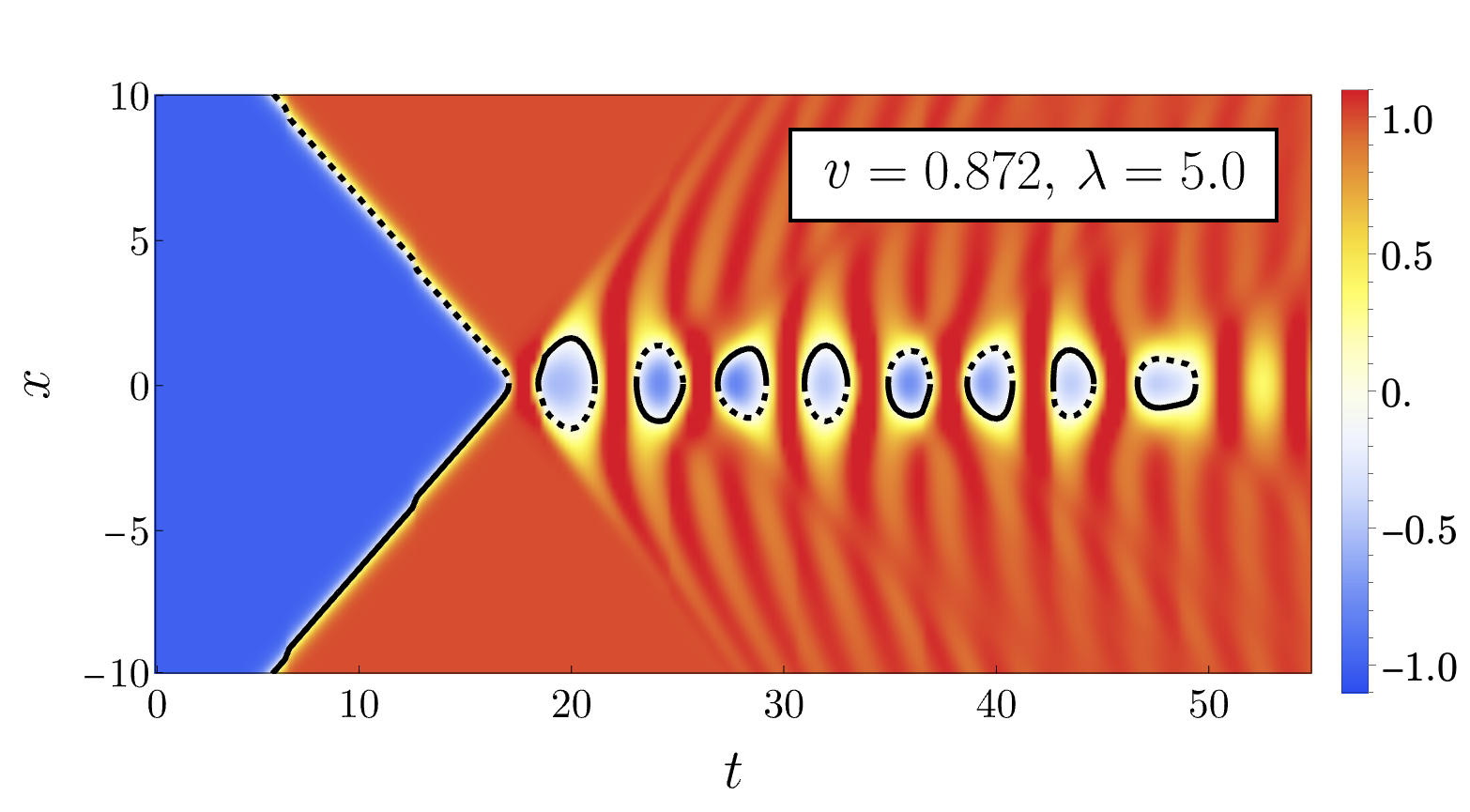}
    \includegraphics[trim=80 66 80 30,clip,width=0.301\textwidth]
    {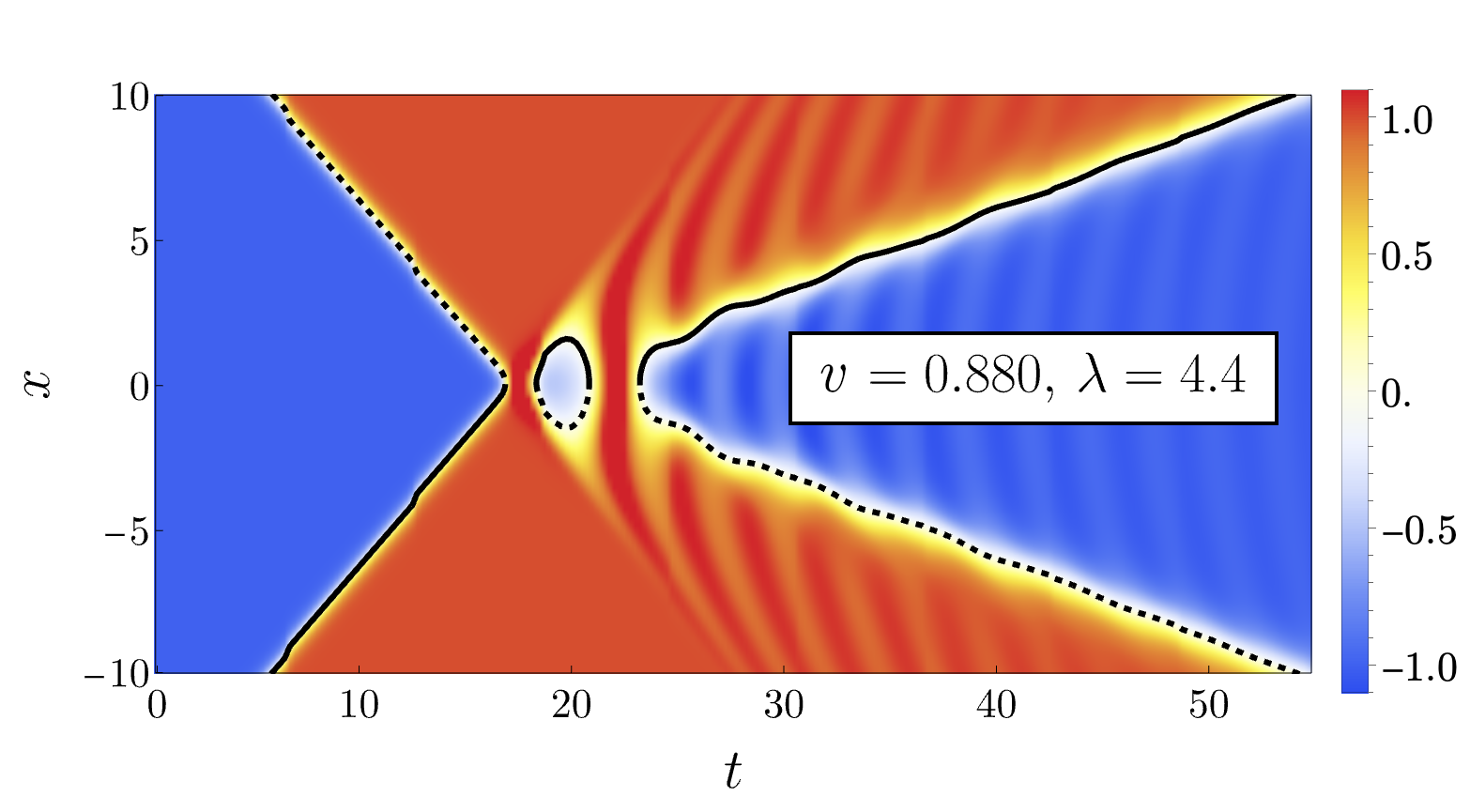}
    \includegraphics[trim=80 66 0 30,clip,width=0.34\textwidth]{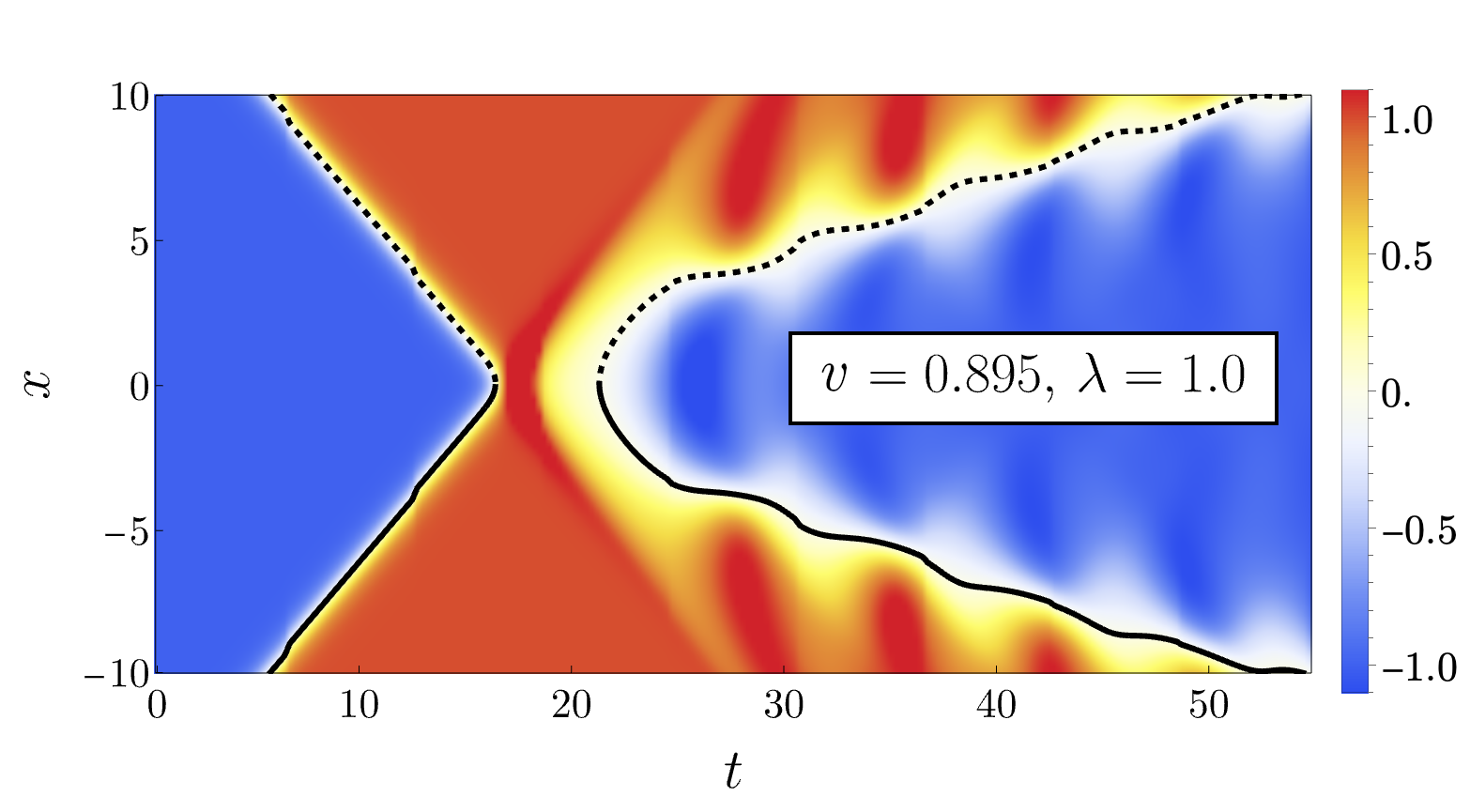}
    \includegraphics[trim=0 0 80 30,clip,width=0.34\textwidth]{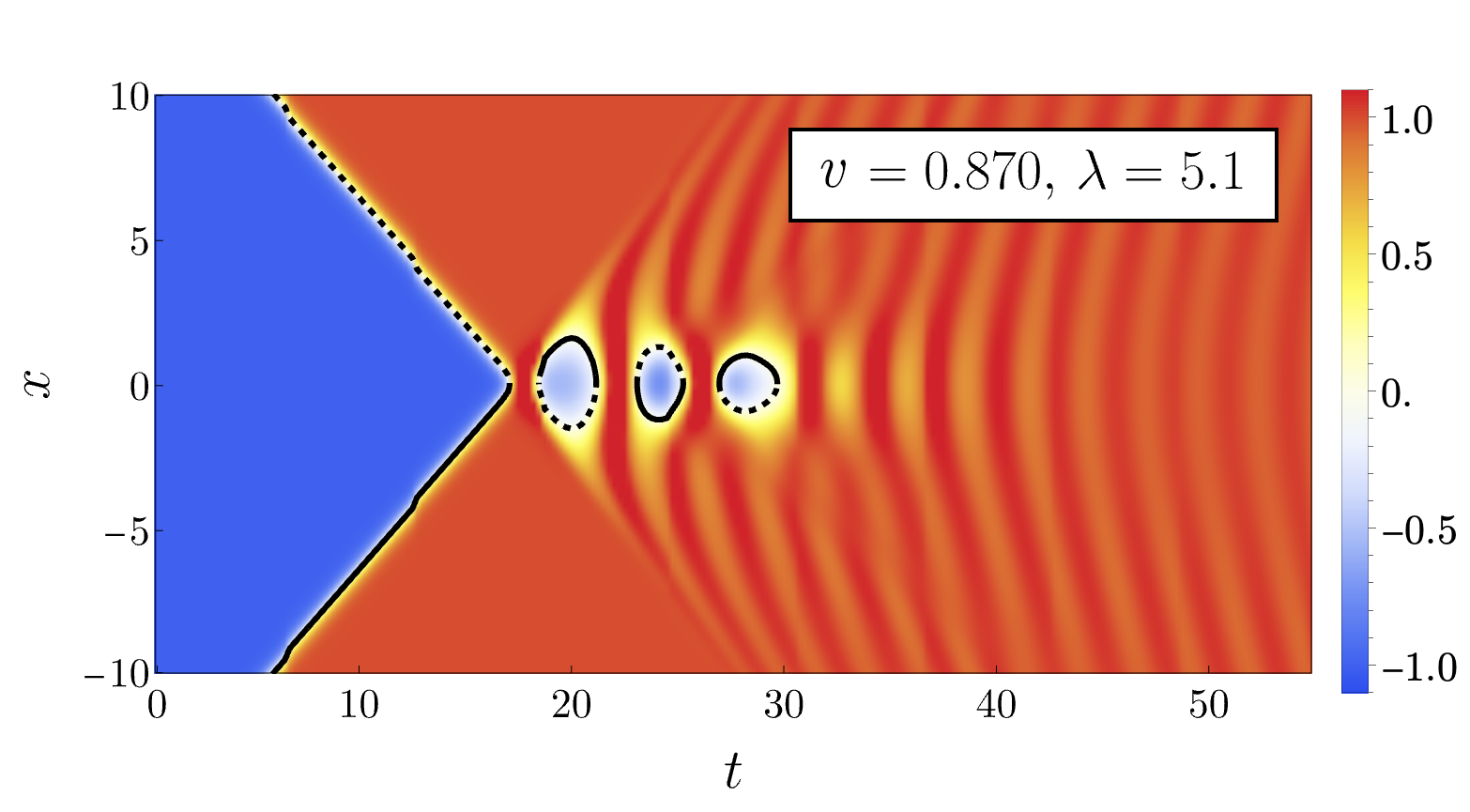}
    \includegraphics[trim=80 0 80 30,clip,width=0.301\textwidth]{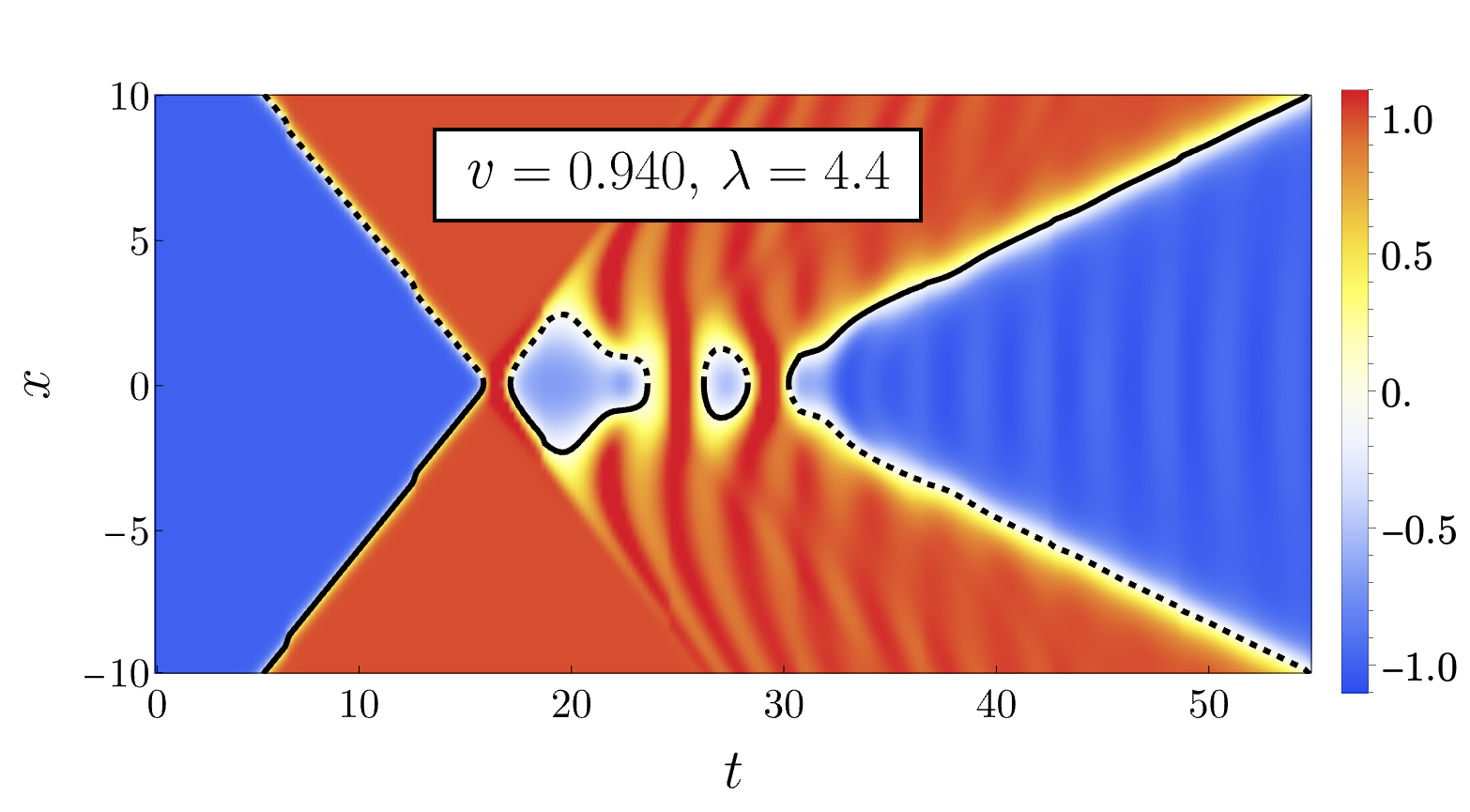}
    \includegraphics[trim=80 0 0 30,clip,width=0.34\textwidth]
    {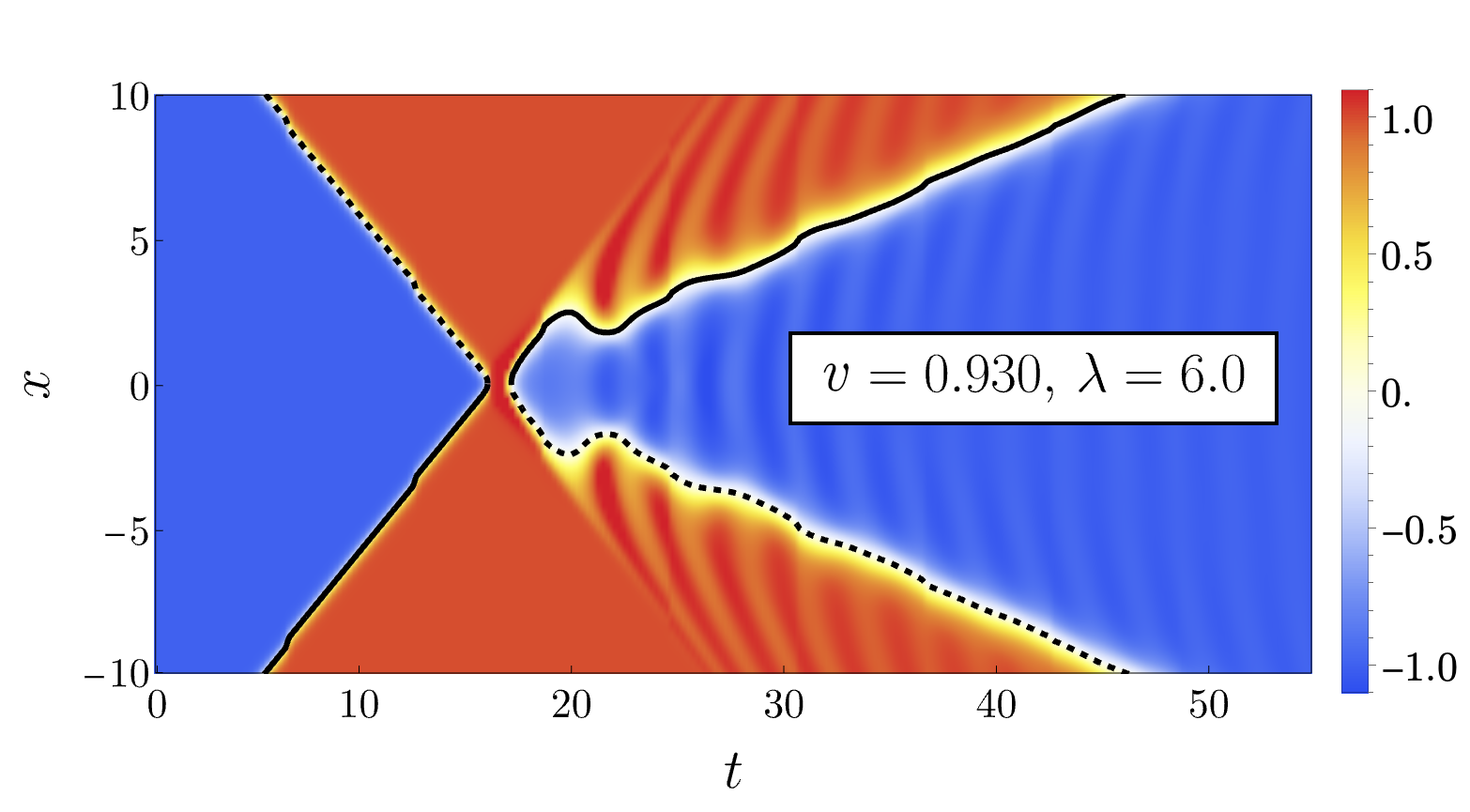}
    \caption{Several examples of vortex-antivortex scatterings are shown. The density plot represents the scalar field $\Re{\phi}$ along the $x$-axis, while the solid (dashed) lines indicate the zeros of the vortex (antivortex).}
    \label{fig:bounce_examples}
\end{figure*}

In Figure~\ref{fig:bounce_examples}, we show particular examples of the scatterings. We plot $\Re (\phi)$ together with the position of the zero of the vortex (solid) and the antivortex (dashed). Interestingly, the temporary recreated solitons, both in the annihilation chimneys and in the bounce windows, have a rather chaotic location, where the backscattering and passing through scenarios happen in an unpredictable way.
A deeper understanding of this phenomenon will
require a more detailed investigation with significantly higher precision.

It is worth noting that the observed behavior, and in particularly Figure~\ref{fig:fractal-structure-lambda44}, resembles the outcome of kink-antikink~\cite{Sugiyama:1979mi, Campbell:1983xu, Manton:2021ipk}, \cite{Gani:2017yla, AlonsoIzquierdo:2020hyp, Campos:2021mkn, Campos:2023xxf, Marjaneh:2023dhu, Karpisek:2024zdj}, two oscillons~\cite{Blaschke:2024uec} or the Q-ball--anti-Q-ball collisions~\cite{Martinez:2025ana}, provided that these solitons carry an internal mode participating in the resonant energy transfer. However, here, for $\lambda>1.5$, where multi-bounce windows occur, the unit charge vortex does not host any bound mode~\cite{Alonso-Izquierdo:2024bzy}. 

\section{The perturbation of the vortex}
\label{sec:the-perturbation-of-the-vortex}

To explain this apparent paradox, let us focus closely on the perturbation around the unit vortex solution:
\begin{align}
    \phi&=f(r)\, e^{i\theta}+u(r)\, e^{i\theta}\, e^{i\omega t},\\
    A_\theta&=\frac{a(r)}{r}-v(r)\, e^{i\omega t},
\end{align}
where $v(r)$ and $u(r)$ are profile functions of the vortex fluctuations. 
The linearized equations (in the background gauge~\cite{Cheng:1984vwu}) for the linear modes are given by
\begin{align}
 &   -\frac{\dd^2u}{\dd r^2} - \frac{1}{r} \frac{\dd u}{\dd r} + \nonumber \\
 & + \left(\frac{(1-a)^2}{r^2} +\frac{3\lambda}{2} f^2 -\frac{\lambda}{2}\right) u +\frac{2}{r} (1-a)f v=\omega^2 u ,
 \label{fluctuation-2}\\
  &  -\frac{\dd^2v}{\dd r^2} - \frac{1}{r} \frac{\dd v}{\dd r} +\left(\frac{1}{r^2} +f^2\right) v +\frac{2}{r} (1-a)f u =\omega^2 v. 
  \label{fluctuation-1}
\end{align}

For $\lambda < 1.5$, the unit charge (anti)vortex has one bound mode \cite{Alonso-Izquierdo:2024bzy}. Of course, its frequency is always below the mass of the Higgs field $m_h=\sqrt{\lambda}$ and the mass of the gauge field $m_v=1$. As $\lambda \to 1.5$, the frequency of the bound mode approaches the mass of the gauge field. For $\lambda > 1.5$, no genuine bound mode remains. However, when crossing the mass threshold, a bound mode does not simply vanish, it typically transmutes into an antibound or quasinormal mode \cite{Alonso-Izquierdo:2024tjc}. In this case, it becomes a Feshbach resonance \cite{Feshbach}, which is a half-bound state in which one component of the perturbation (that of the matter field) remains localized around the vortex, while the other (that of the gauge field) forms a scattering solution.

\begin{figure}
    \centering
    \includegraphics[trim=0 0 0 25, clip, width=\linewidth]{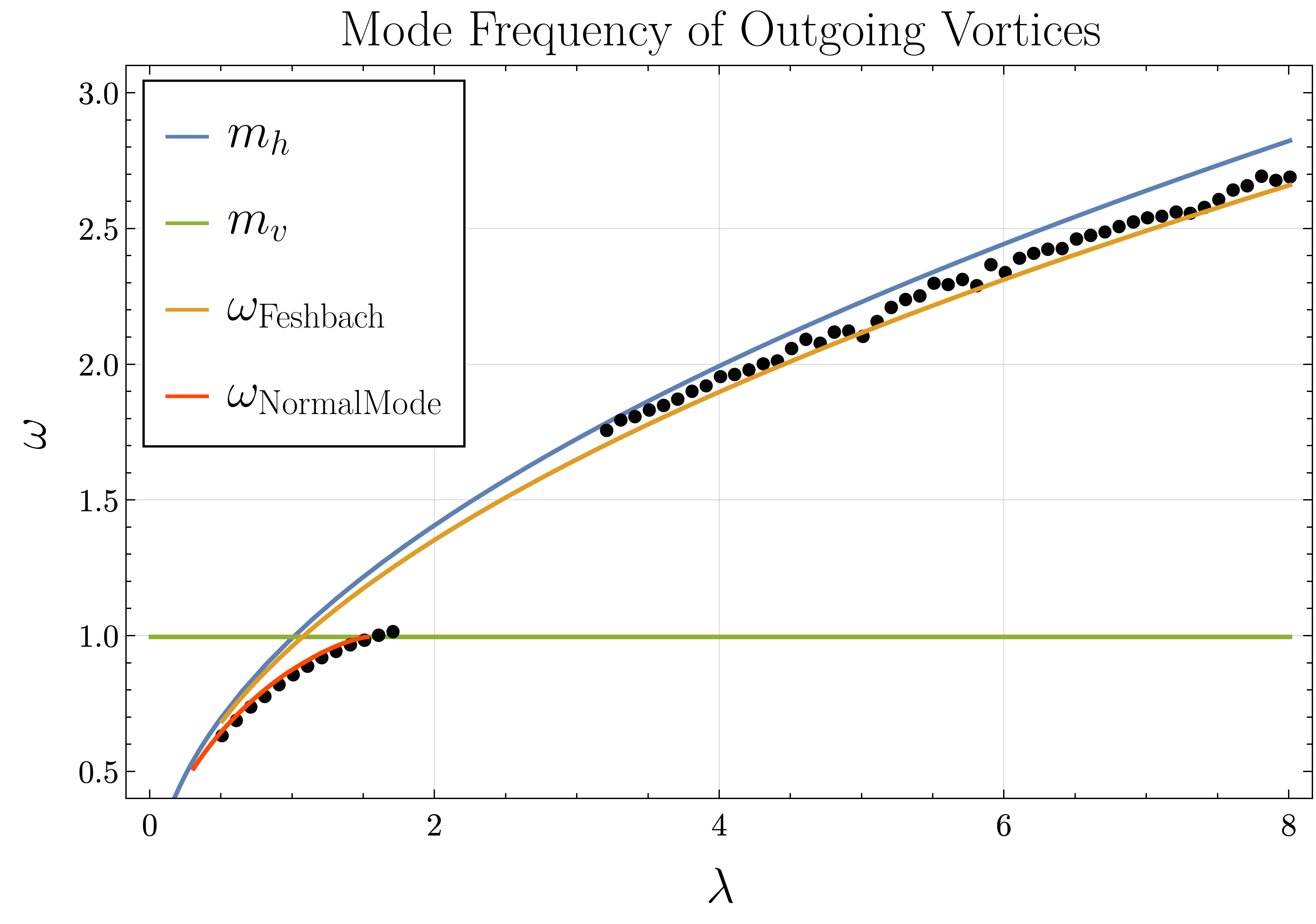}
    \caption{Measured frequency of the vibrating vortex recreated in one-bounce collisions (black dots) vs. the approximated frequency of the Feshbach resonance (orange curve) and the bound mode (red curve). Blue and green curves are the mass thresholds of the Higgs and gauge fields respectively.}
    \label{fig:modes}
\end{figure}

To find the Feshbach resonance, in an approximate way, we assume the decoupled version of the problem, where only the matter field has a nontrivial perturbation (thus, $v(r)\equiv0$). This reflects the fact that the matter field component would have a bound mode localized on the vortex.
Furthermore, we suppress the off-diagonal terms in the fluctuation problem~\eqref{fluctuation-2}, \eqref{fluctuation-1}, which are effectively responsible for the decay of the perturbation into gauge radiation (see Supplemental Material). This leaves us with a single equation for $u$. The frequency of our approximation to the lowest Feshbach mode is shown in Figure~\ref{fig:modes} (orange curve).

To confirm that this is indeed this mode, which participates in the resonant energy transfer mechanism and triggers the multi-bounce structure in the final-state formation, we measure the frequency of the vibrations of the outgoing vortex, see Figure~\ref{fig:modes} (black dots). The outgoing vortex vibrates for $\lambda \gtrsim 3.2$ and the measured frequency agrees very well with the frequency of the Feshbach mode. Hence, this is indeed the mode that is excited during the scattering and is responsible for the multi-bouncing behavior.

The general feature of this decoupled problem is the growing number of half-bound modes as $\lambda$ increases. Asymptotically, for $\lambda \to \infty$, we recover the global vortex model, which is known to support infinitely many bound modes in the unit vortex sector~\cite{Blanco-Pillado:2021jad}. The growing number of Feshbach resonances may be a reason why the multi-bounce structure is not observed for large $\lambda$. Indeed, in this case, it is much less probable that energy transferred into the internal modes may at some point flow back to the kinetic DoF, which would allow for the reappearance of the vortices in the final state (see analogous behavior in kink-antikink collisions \cite{Simas:2016hoo}). 

For $\lambda < 1.5$ the frequency of the vibration of the outgoing vortex is just the frequency of its bound mode.

\section{Conclusions and Outlook}
In the present work, we confirm that the resonance energy transfer mechanism is a phenomenon that is not restricted to one spatial dimension and governs the scattering of vortices qualitatively in the same fashion as in the case of domain walls. In the deep type II regime, we clearly established the existence of a non-trivial pattern in the final state formation with multi-bounce windows chaotically immersed in annihilation regions. Surprisingly, the resonant energy transfer mechanism is triggered by the lowest Feshbach resonance existing in the deep type II regime. It is also surprising that for the investigated parameter space we haven't found such pattern for $\lambda<1.5$, where the genuine shape mode of the vortex exists and should potentially enter the resonant energy transfer. We note that we did not find flux-swapping oscillons~\cite{Gleiser:2007te} in the annihilation scenarios. This is consistent with the fact that in the AH model, oscillons exist only for small $\lambda$~\cite{Gleiser:2007te}.   

The observed complicated structures in the vortex-antivortex collisions may be important for understanding the dynamics of cosmic strings~\cite{Vilenkin:2000jqa, Hindmarsh:1994re}, see full field theory computation~\cite{Vincent:1997cx, Moore:1998gp, Helfer:2018qgv, Correia:2020yqg} or the Nambu-Goto approximation~\cite{Albrecht:1984xv, Bennett:1989ak, Allen:1990tv, Martins:2005es, Ringeval:2005kr, Blanco-Pillado:2011egf, Blanco-Pillado:2013qja}. 

There are several ways in which the current work can continue. First, it would be very desirable to reproduce the full field theory scatterings in terms of a collective coordinate model, exactly as was done for the kink-antikink collisions in the $\phi^4$ model~\cite{Manton:2021ipk}.
For example, in~\cite{Bradshaw} the authors construct an iterative model which captures the dynamics of wobbling kinks based on quantitative features of the underlying field theory. This model shows evidence of universality, which suggests that it should be applicable to vortex-antivortex scattering.
However, since the vortex collisions occur in a strongly relativistic regime, it will be important to take into account the deformation of the moduli space metric by the amplitude of the internal mode~\cite{Miguelez-Caballero:2025xfq}. 

Second, since the Feshbach resonance(s) take over the role of the bound mode for $\lambda>1.5$, one can expect that vortex-vortex collisions in the deep type II regime may also reveal a non-trivial pattern. In particular, the symmetric (in-phase) excitation of the Feshbach resonance on each of the vortices may lead to a temporary bound state of two vortices. 

Undoubtedly, a detailed study of the quasinormal modes, including the Feshbach resonances, of the vortices in the Abelian-Higgs model is very desired. As a first step, this can be addressed in a perturbative way along the line considered in~\cite{GarciaMartin-Caro:2025zkc} and motivated by~\cite{Manton_1997}. This should allow for the computation of the decay rate of the Feshbach resonances. A more direct study would require a consideration of the coupled linear perturbation problem with the purely outgoing boundary condition, see e.g.,~\cite{Adam:2017czk} and~\cite{Dorey:2017dsn}.

Third, it would be interesting to analyze the vortex-antivortex process in the case of excited vortices. In fact, while produced in a phase transition, vortices typically carry excitations of the internal modes. Such modes can also be excited by an oscillating background, e.g., provided by an inflaton~\cite{Kitajima:2025nml}.

Fourth, a more detailed study is required to examine the effect of introducing a small impact factor, implemented as a slight shift perpendicular to the direction of motion. First simulations show that, for small impact factors, the vortices can still bounce and recreate. However, the outcome can change significantly, and thus all windows in Figure~\ref{fig:chaoticmap} can be strongly deformed (for instance, for $\lambda=1$, velocity $0.9$, and an impact factor of $\Delta y=1$, the vortices finally annihilate instead of backscatter, as in the case without an impact factor). One reason for this drastic change is an additional excitation that appears due to the nonzero initial angular momentum. During the interaction, part of the angular momentum is transferred into an internal excitation of the vortices that carries angular momentum. This effect is strong for small impact factors, becomes weaker for larger impact factors, and appears only when the vortex cores overlap. For very large impact factors, this angular momentum transfer does not occur, and the vortices pass by each other, changing their direction of motion only slightly due to the attractive force between a vortex and an antivortex. This angular-momentum-carrying excitation of charge-1 vortices is a novel type of excitation. Its exact form and implications for vortex-antivortex collisions will be investigated in a future study.

Of course, one may also wonder whether the scattering of the vortices in various extensions of the Abelian-Higgs model reveals a similar chaotic behavior, see e.g., \cite{Kim:2025ien, Kim:2025mml, Leask:2025vgx, Kitajima:2025jct}. The same question can be asked for non-Abelian vortices~\cite{Eto:2011pj, Eto:2020anv}.

Looking from a more general perspective, the universality of the resonant energy transfer mechanism implies that very similar results can be expected in the case of monopole-antimonopole collisions, see~\cite{Vachaspati:2015ahr} for annihilation of the 't Hooft-Polyakov monopole-antimonopole pair. Importantly, it is known that Feshbach resonances are internal modes of the BPS monopole~\cite{Fodor:2003yg, Forgacs:2003yh, Fodor:2006ue, Russell:2010xx}. Thus, our result gives further evidence that the interaction of excited BPS monopoles should go much beyond the geodesic approximation \cite{Manton:1981mp,Atiyah:1985dv, Atiyah:1985fd, Gibbons:1986df,Bachmaier:2025jaz} and may reveal an analogous chaotic structure.  

\section*{Acknowledgement}
The authors thank A. Alonso Izquierdo, G. Dvali, S. Krusch, N. Manton, and T. Romanczukiewicz for discussion and remarks. Moreover, the authors thank A. Alonso Izquierdo for sharing the numerical data concerning the bound mode of the unit vortex.

\section*{Supplemental Material}
\subsection*{\normalsize Appendix A: Numerical implementation}
\textbf{Initial configuration.}
As an initial ansatz for a colliding vortex-antivortex pair, we take the product ansatz 
\begin{align}
    \phi_\text{v-av}(t,x,y)=\, &\phi_+(\gamma_1 (x-v_1 t-d/2),y)\nonumber\\
    \cdot\, &\phi_-(\gamma_2 (x-v_2 t+d/2),y),
\end{align}
where $\phi_\pm$ are the scalar fields for a single vortex/antivortex given in equation~\eqref{eq:vortex-ansatz}. $d$ is the separation distance between the two vortices. $v_1$, $v_2$ are the velocities of the vortex and antivortex respectively and $\gamma_1$, $\gamma_2$ are the corresponding Lorentz factors.

Since the theory is Higgsed, the gauge field is massive and consequently of short range outside of the vortex core. This allows us to simply add the gauge fields corresponding to the vortex and antivortex:
\begin{align}
    A_\mu(t,x,y)=&\begin{pmatrix}
        -v_1 \gamma_1 A^+_x(\gamma_1(x-v_1 t-d/2),y)\\
        \gamma_1 A^+_x(\gamma_1(x-v_1 t-d/2),y)\\
        A^+_y(x,y)
    \end{pmatrix}\nonumber\\
    +&\begin{pmatrix}
        -v_2 \gamma_2 A^-_x(\gamma_2(x-v_2 t+d/2),y)\\
        \gamma_2 A^-_x(\gamma_2(x-v_2 t+d/2),y)\\
        A^-_y(x,y)
    \end{pmatrix}.
\end{align}

The separation distance, for the plots presented in this work, was chosen to be $d = 30$. For the creation of Figure~\ref{fig:chaoticmap}, we scanned the parameter space within the intervals $\lambda \in [0.1,8]$ with a step size of $\Delta \lambda = 0.1$ and $\abs{v_1} = \abs{v_2} \in [0.8, 0.98]$ with a step size of $\Delta v_1 = \Delta v_2 = 0.01$. For the cases $\lambda = 4.4$ and $\lambda = 4.9$, shown in Figures~\ref{fig:fractal-structure-lambda44} and~\ref{fig:fractal-structure-lambda49}, respectively, a finer precision of $\Delta v_1 = \Delta v_2 = 0.001$ was used.\\

\textbf{Numerical methods.}
For the time integration, we used the second order Runge-Kutta method (RK2) (for more details see~\cite{Figueroa:2020rrl}).
We used natural boundaries as explained in~\cite{Krusch:2024vuy} in combination with the adiabatic damping method~\cite{Gleiser:1999tj}, that is, outside a region of radius $25$, we added a friction term to the field equations with a prefactor that increases as a Gaussian function for larger radii.
This boundary condition is perfectly suited for our setup, since even for strong Lorentz boosts the radiation either emitted from or reflected at the boundary remains negligibly small, ensuring that it doesn't influence the scattering dynamics by a lot.

Furthermore, the lattice size was chosen sufficiently large to ensure that any radiation emitted from the boundary initially doesn't reach the origin (collision point) before the vortices collide.

The simulations were carried out on a square lattice with $x, y \in [-30, 30]$ and a spatial spacing of $0.05$. The investigated time interval was chosen to be $[0, 100]$ with a time step size of $0.02$. For numerical tests, the lattice size and resolution was varied showing no significant change in the results.\\

\textbf{Programming language.}
For the numerical simulations we used the programming language Python together with the packages Numpy and Numba~\cite{Numba}. Numba translates the Python code into a fast machine code and provides a straightforward way for parallelizing the code. Therefore, it enhances the computation speed by a lot. The figures were created using Mathematica.

\subsection*{\normalsize Appendix B: Frequency and velocity measurement}

As discussed in section~\ref{sec:the-perturbation-of-the-vortex}, the vortices recreated after the collision carry excited modes. These fluctuations can also be clearly seen in Figure~\ref{fig:bounce_examples}. For $\lambda \leq 1.5$, the excited mode corresponds to the bound mode, whereas for $\lambda \gtrsim 3.2$, it corresponds to the lowest Feshbach mode.

From our numerical simulation data, we extracted the frequency of the mode oscillations by integrating the total potential energy of the system. Although the total energy is conserved, the potential energy is not. 
There is permanent flow between the kinetic and potential energy due to the exited modes. This results in oscillations in the potential energy and provides a clear observable from which the frequency can be determined.

Due to the high velocities of the recreated vortices and the comparably small lattice size, the number of observable mode oscillations is insufficient for a fully precise Fourier analysis. Therefore, we measured the time interval over which the oscillations are clearly visible and divided it by the number of oscillations within this interval to obtain the period, from which the frequency can be determined. The results are given in Figure~\ref{fig:modes}.\\

\begin{figure}[t]
    \centering
    \includegraphics[trim=0 0 0 38, clip,width=\linewidth]{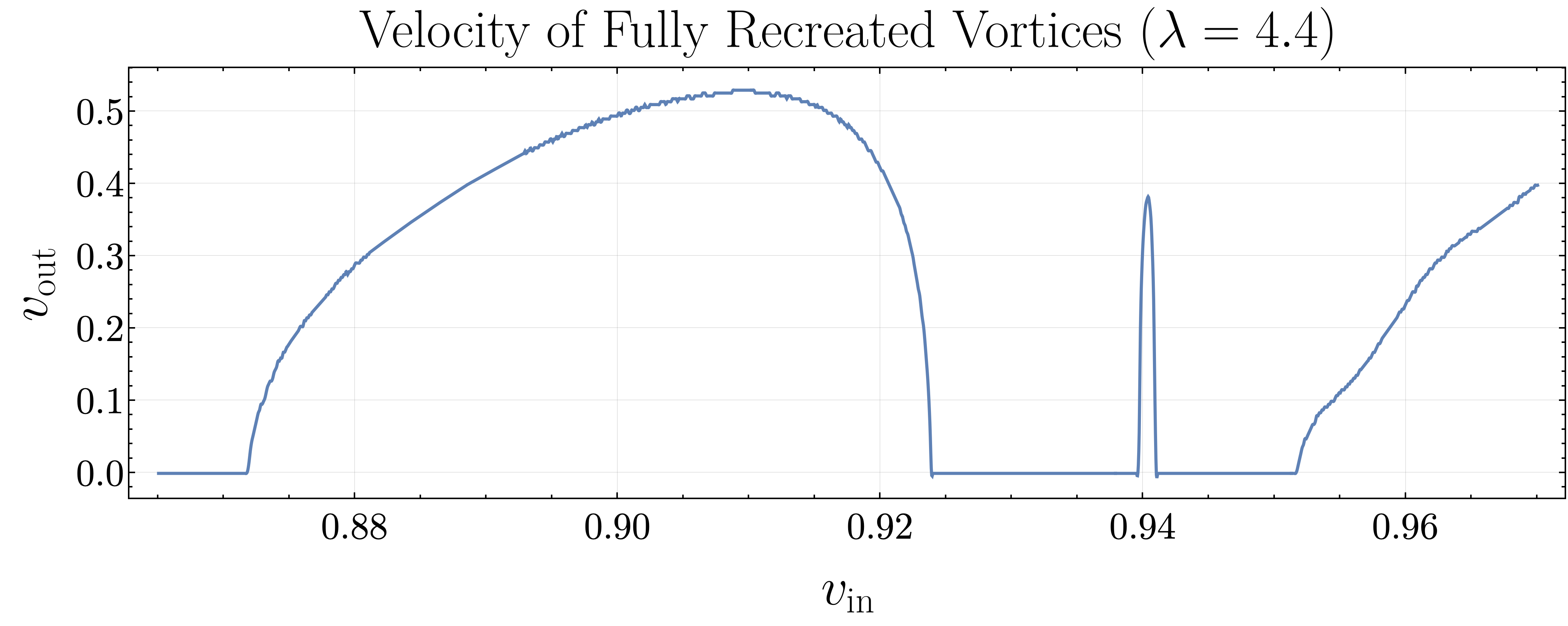}
    \caption{VAV scattering for $\lambda=4.4$. Final velocity of recreated vortex as a function of $v_{in}$. }
    \label{fig:outgoingvelocity}
\end{figure}

\begin{figure}[t]
    \centering
    \includegraphics[trim=0 0 0 35,clip,width=\linewidth]{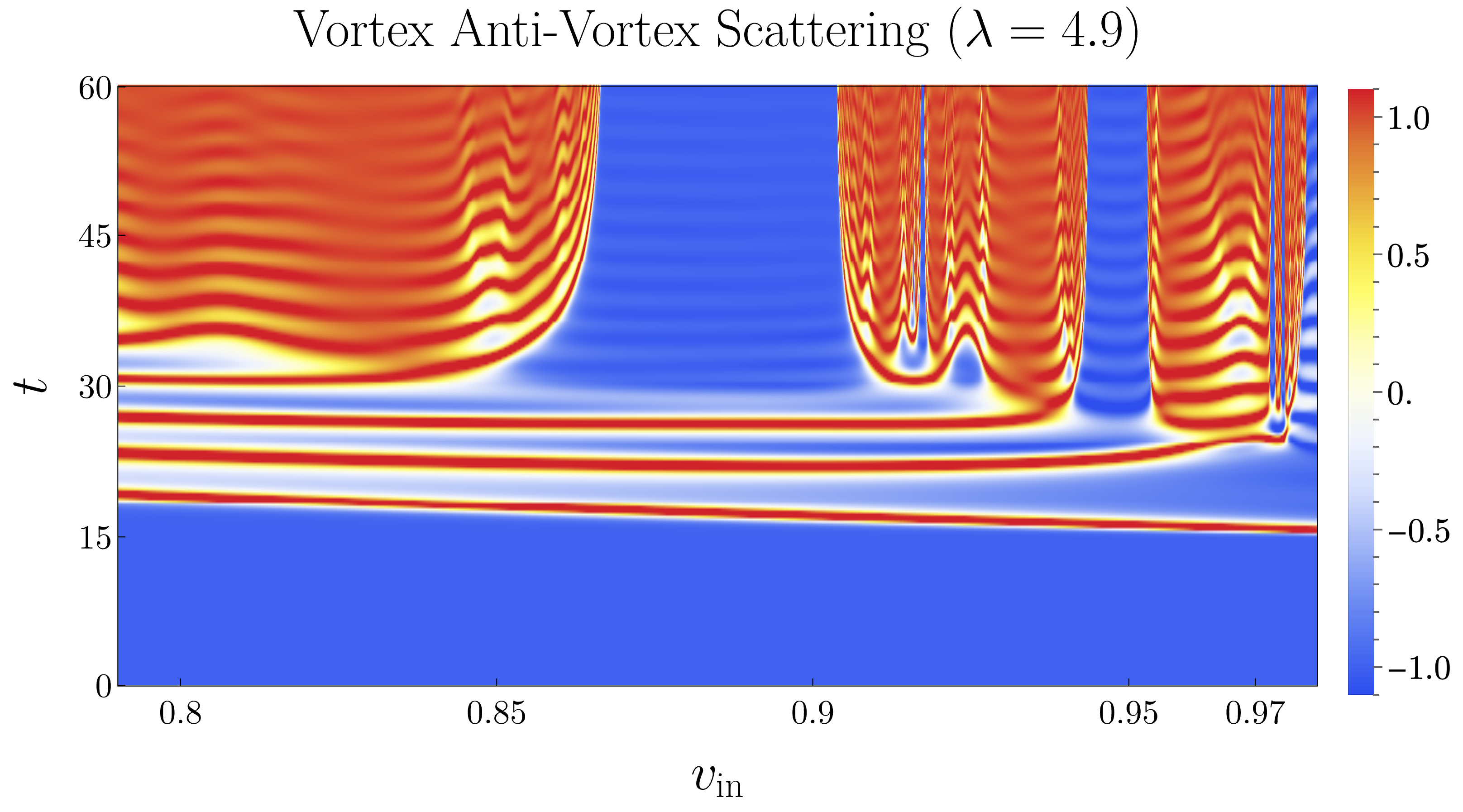}
    \caption{Time evolution of the real part of the scalar field at the origin, $\Re \phi(\vec{x}=0, t)$, during a vortex-antivortex scattering for $\lambda = 4.9$, shown for different initial velocities $v_{\rm in}$.}
    \label{fig:fractal-structure-lambda49}
\end{figure}

For $\lambda = 4.4$ and $\lambda = 4.9$, we analyzed vortex-antivortex collisions with a finer velocity spacing ($\Delta v_1 = \Delta v_2 = 0.001$). This higher resolution allowed us to obtain a clearer picture of the recreation and multi-bounce windows (see Figures~\ref{fig:fractal-structure-lambda44} and~\ref{fig:fractal-structure-lambda49}). The scattering behavior for $\lambda = 4.4$ is very similar to that observed in one-dimensional scenarios such as kink-antikink, two-oscillon, or Q-ball-anti-Q-ball collisions.
The velocity of the zeros after vortex recreation within the recreation regime was measured to determine the dependence of the outgoing velocity on the initial velocity. The corresponding results are shown in Figure~\ref{fig:outgoingvelocity}. 

We observe that within a recreation window, the outgoing velocity increases with the initial velocity, reaches a local maximum, and then decreases again. Above the final recreation velocity, the outgoing velocity becomes a monotonically increasing function.
Furthermore, in the investigated parameter regime, the outgoing velocity never reaches values comparable to the initial velocity, indicating that a significant fraction of the initial kinetic energy is converted into mode excitations and radiation emitted during the collision.
Overall, this behavior closely resembles that found in the well-known examples of  one-dimensional theories.\\

In addition to the $\lambda=4.4$ case, we analyzed the $\lambda=4.9$ case in more detail. This case turned out to be particularly interesting, since the resulting scattering pattern differs significantly from those observed for other values of $\lambda$. The results are given in Figure~\ref{fig:fractal-structure-lambda49}.

First, we can observe two large recreation windows, one with three bounces and another one with two bounces. Such a large interval for three bounces with subsequent recreation is rather unusual compared to the one-dimensional examples, making this case special. Additionally to the two main recreation windows, we can see several smaller recreation regions with up to four bounces. Furthermore, there is also a false window around $v_{\rm in}\approx 0.925$. This is a clear three bounce collision, which eventually results not in reappearance of the vortices in the final state but in their complete annihilation. This is another typical feature of the chaotic pattern of kink-antikink collisions.

Notice that the results for the $\lambda=4.9$ case should be taken with caution. First, the value of $\lambda$ lies in the transition region between the two regimes, one in which the vortices backscatter after the final recreation and another one in which they pass through each other.
A complete understanding of these two behaviors is still waiting to be discovered. Second, the highly relativistic regime for velocities above $v_1=v_2=0.95$ requires a more detailed analysis with higher lattice resolution.

\subsection*{\normalsize Appendix C: Winding measurement}
\begin{figure}[t]
    \centering
    \includegraphics[trim=0 0 0 0,clip,width=0.9\linewidth]{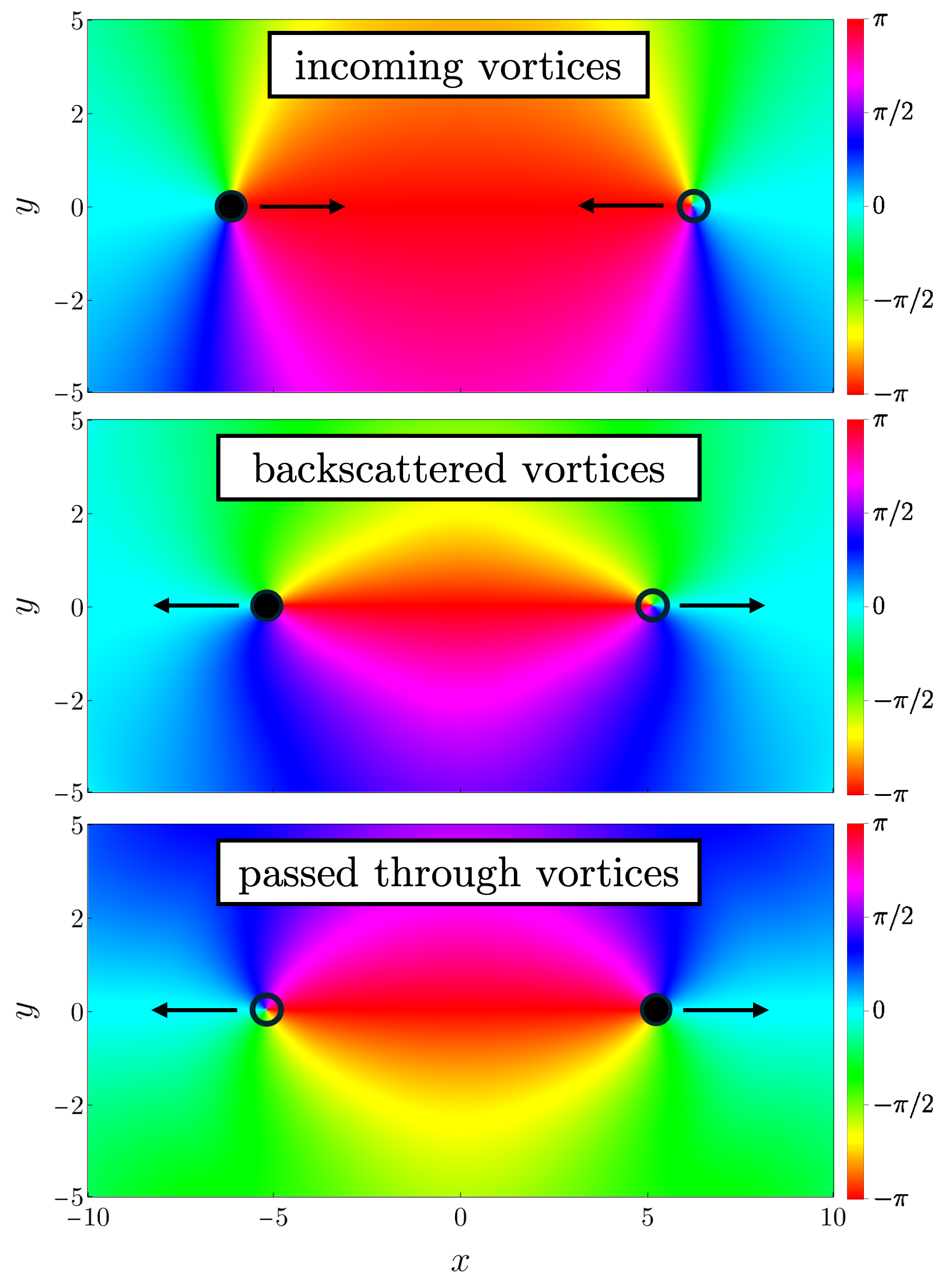}
    \caption{Illustration of the two possible outcomes when the vortex and antivortex are fully recreated. The ``rainbow" plot shows the phase of the complex scalar field $\phi$. Filled (empty) circles indicate the positions of the vortex (antivortex) cores.}
    \label{fig:naming}
\end{figure}

To determine the position of the vortices we tracked the positions of the zeros of the field. Beyond the positions themselves, the winding around each zero is an important information, because it is allowing us to distinguish between vortices and antivortices. The winding was obtained by visualizing the full time evolution of the phase of the complex scalar field $\phi$ using a ``rainbow" plot. The positive winding is identified as anti-clockwise winding of the phase. Here, the phase changes from $-\pi$ to $\pi$, which corresponds to the transition from red to red through yellow, green, blue, and purple. A few illustrative examples are presented in Figure~\ref{fig:naming}.

As mentioned in the main text, when the vortices fully recreate and separate forever, two possible scenarios can occur. In the first, the vortices backscatter ($180^\circ$ scattering), meaning the vortex entering from the left also exits to the left. In the second, they pass through one another ($0^\circ$ scattering), so that the vortex entering from the left exits to the right. Figure~\ref{fig:naming} illustrates our naming convention for these two scenarios.

\subsection*{\normalsize Appendix D: Feshbach resonances}

A Feshbach resonance is a particular type of quasinormal mode. It exists in a multi-channel system provided that, in the decoupled limit, one of the channels possesses bound modes, which, simultaneously lay {\it above} the mass threshold of the other channels. \\

In the simplest case one needs a two-channel problem, which in $2+1$ dimensions take the following form
\begin{align}
   -\frac{1}{r}\frac{d}{dr} \left( r \frac{d u }{dr} \right) + U_u(r)u + \mu W(r) v&=\omega^2_n u, \\
     -\frac{1}{r}\frac{d}{dr} \left( r \frac{dv }{dr} \right) + U_v(r)v + \mu W(r) u&=\omega^2_n v,
\end{align}
where $U_u(r)$, $U_v(r)$, $W(r)$ are potentials (e.g., arising from the unperturbed vortex solution). The off-diagonal part is formally multiplied by a coupling parameter $\mu$. We assume that the problem has two {\it different} mass thresholds $m_u$ and $m_v$
\begin{equation}
    \lim_{r\to \infty} U_{u}= m^2_{u}\, , \hspace{1cm} \lim_{r\to \infty} U_{v}= m^2_{v}\, .
\end{equation}
We also assume that in the decoupled limit, $\mu=0$, the first channel admits positive energy states $u_n$ with $\omega_n^2<m_u^2$. In addition, these modes should be {\it above} the mass threshold of the second channel, $\omega_n^2 > m^2_v$. Thus, they are located in the continuum spectrum of this channel. As there is no coupling between the channels (at the quadratic order), these are genuine bound modes located on the analyzed solution. 

However, once we switch on the off-diagonal part, the localized perturbation, that is, the bound mode, begins to excite the second channel. However, since the bound mode is above the mass threshold in the second channel, this results in an excitation of scattering states. Eventually, this leads to the decay of the mode even at the quadratic level. Qualitatively, the decay rate depends on the off-diagonal mixing potential and the mass threshold of the second channel. 

In our work, we consider the decoupled version of the linear problem and test the validity of this assumption {\it a posteriori}. Namely, we compare the frequency found in the decoupled linear problem with the actual measured frequency of the vortex. The agreement is very good, suggesting that the off-diagonal terms can be neglected, at least at the leading approximation, and as long as the real frequency of the oscillation is considered.

Importantly, as $\lambda \to \infty$ we tend to the global string case. Notice that this limit is usually achieved by taking $g\to 0$. Now, the gauge sector completely decouples, and the unit vortex has infinitely many proper discrete bound modes. For finite but large $\lambda$ the Higgs and the gauge sectors are coupled with some inverse power of $\lambda$, and it is plausible to expect that the lowest modes are still very narrow resonances, which means a small energy transfer between the sectors. Of course, this results in a (very) small decay constant, which makes these modes relevant for the resonant energy transfer mechanism. This also agrees with our numerical analysis, where the vortices host long-lived oscillations only for larger $\lambda$. All this further supports our approximate treatment of the Feshbach resonances.


\bibliography{ref}

\end{document}